\documentclass[twocolumn,showpacs,preprintnumbers,amsmath,amssymb]{revtex4}
\usepackage{graphicx}
\usepackage{amsfonts}
\usepackage{amssymb}
\usepackage{bm}
\usepackage{psfrag}
\usepackage{subfigure}
\usepackage{graphicx}
\usepackage{rotating}
\usepackage{color}
\def\strutdepth{\dp\strutbox}
\def\nw#1{\strut\vadjust{\kern-\strutdepth\vtop to0pt{\vss\hbox to\hsize
{\hskip\hsize\hskip5pt$\leftarrow$\hss\strut}}}{\em #1}}

\begin{document}

\title{ Comment on ``Force Balance at the Transition from Selective 
Withdrawal to Viscous Entrainment'' }

\author{J. Eggers\dag\ and S. Courrech du Pont\ddag  }

\affiliation{
$^\dag$School of Mathematics, 
University of Bristol, University Walk, 
Bristol BS8 1TW, United Kingdom  \\
$^\ddag$ Laboratoire Mati{\`e}re et Syst{\`e}mes Complexes, 
UMR CNRS 7057, Universit{\'e} Paris Diderot \\
10, rue Alice Domon et L{\'e}onie Duquet, 75205 Paris cedex 13, France
                  }
\maketitle

\begin{figure}
\begin{center}
\psfrag{S}{$S/\ell_c$}
\psfrag{ss}{$S$}
\psfrag{Q}{\begin{rotate}{90}$Q_c (\mu/\ell_c^2\sigma)$\end{rotate}}
\psfrag{r}{$r/\ell_c$}
\psfrag{z}{$\delta z$}
\psfrag{m}{$\mu$}
\psfrag{s}{$\sigma$}
\psfrag{m0}{$\textcolor{white}{\mu_0}$}
\psfrag{f}{fluid}
\psfrag{a}{\textcolor{white}{air}}
\psfrag{o}{sink}
\includegraphics[scale=0.5]{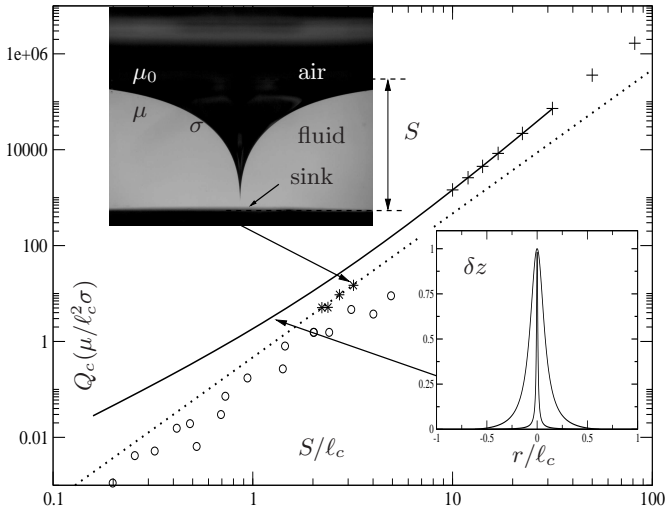}
\caption{\label{bound} 
Critical flow rate versus distance between sink and the unperturbed 
interface $S$, in units of the capillary length $\ell_c$.
The line is our numerical simulation for $\mu_0/\mu=1$, o:
data from \cite{C04} as shown in \cite{BZ09}, 
$*$: data from \cite{CE06}; $Q$ multiplied by 3 \cite{CE06},  
$+$: simulation from \cite{L89}. Dotted line:
$Q_c = 2.2\gamma S^3 / \mu \ell_c$. Top inset: image of fluid-air 
interface as the tip enters the tube, \cite{CE06}. Bottom inset: failure
mode $\delta z$ for $S/\ell_c=1.3$ and $\mu_0/\mu=1$ as well as 
$\mu_0/\mu=0.1$ (narrow peak). 
               }
\end{center}
\end{figure}

In  a recent letter \cite{BZ09}, Blanchette and Zhang (BZ) proposed
a theory for the critical flow rate at the selective withdrawal 
transition. Their theory is based on the assumption of failure 
of the interface on a large scale, presumed insensitive to the viscosity 
$\mu_0$ of the entrained fluid. We show that BZ's theory is untenable, 
as it is inconsistent with the hydrodynamic description they use, 
and also disagrees with experiments done in a fluid-air system 
in the same geometry \cite{CE06}, in which no failure was observed, 
and entrainment only occurs when the highly deformed tip of the interface 
enters the orifice, see Fig.~\ref{bound}, top inset.

We performed our own numerical simulations of the hydrodynamic equations,
reported in Fig.~\ref{bound}. We show the critical flow rate at a 
viscosity ratio $\mu_0/\mu=1$ (solid line), using an 
infinite domain (for details, see \cite{EC09}), as appropriate for the 
very large tank used in the two-fluid experiment \cite{C04}. Our results are 
in perfect agreement with earlier numerical calculations \cite{L89} for 
large $S/\ell_c$ (+). In the lower inset we show the (linearly) 
unstable mode at failure for $S/\ell_c = 1.3$, leading to entrainment. 
Contrary to BZ's claim, the unstable mode is highly localized near 
the central peak, and much more so as $\mu_0/\mu$ is decreased to 1/10. 
Thus viscous hydrodynamics, used as the basis for BZ's argument, predicts
an entrainment mechanism inconsistent with their central assumption of 
delocalized failure, nor can the viscosity ratio be neglected, as the
most unstable mode depends strongly on it.

We now discuss our experiments in a fluid-air system \cite{CE06}, in 
which {\it no} instability was observed, and entrainment occurs
only when the hump tip enters the tube. If the interior viscosity
$\mu_0$ may be neglected, the interface shape is well described by 
Taylor's theory \cite{EC09}. Postulating that the deformation occurs 
over a distance $r\propto\ell_c$, one finds for the flow rate as 
the hump height is maximum ($H=S$) that  $Q_c = a\gamma S^3 / \mu \ell_c$. 
This law is shown as the dotted line 
in Fig.~\ref{bound}, with $a=2.2$, which agrees very well with our 
experiment for the fluid-air system (*). One can see in Fig. 2 of 
BZ that the transition occurs in a parameter range where the hump height $H$
is very sensitive to small variations of $Q$, so that the $S$-value
at the transition is close to the marginal case $H=S$ (realized in the 
fluid-air experiments). This explains 
the good alignment of the data from \cite{CN02} with $S^3$ scaling.

Notice that the solid line lies considerably {\it above} the critical 
flow rates of the fluid-fluid experiment (o),
and even above the fluid-air data (*). We believe the latter to be due
to finite size effects, which deform the interface away from
the tip (see Fig. 17 of \cite{EC09}). As for the fluid-fluid data,
taken in a very large tank, it remains unclear which experimental feature
is missing from the hydrodynamic description, as discussed in detail 
in \cite{EC09}. We believe the agreement between theory and experiment
reported by BZ to be an artefact of the unphysical boundary condition
imposed at the edge of a very small domain, chosen arbitrarily to be of
size $\ell_c$. This amounts to an adjustable parameter, and thus an 
arbitrary shift in the $x$-direction. In conclusion, the arguments of 
BZ do not explain the experimental data.

\end{document}